%% file: main.tex
\documentclass[sigconf]{acmart}

\AtBeginDocument{%
  \providecommand\BibTeX{{%
    \normalfont B\kern-0.5em{\scshape i\kern-0.25em b}\kern-0.8em\TeX}}}

\copyrightyear{2025}
\acmYear{2025}
\setcopyright{rightsretained}
\acmConference[CHI EA '25]{Extended Abstracts of the CHI Conference on Human Factors in Computing Systems}{April 26-May 1, 2025}{Yokohama, Japan}
\acmBooktitle{Extended Abstracts of the CHI Conference on Human Factors in Computing Systems (CHI EA '25), April 26-May 1, 2025, Yokohama, Japan}\acmDOI{10.1145/3706599.3720033}
\acmISBN{979-8-4007-1395-8/2025/04}

\usepackage{subcaption}
\usepackage{makecell}
\usepackage{colortbl}
\usepackage{tabularx}

\definecolor{LightGray}{rgb}{0.929,0.929,0.929}

\begin{document}

\title{AI Literacy Education for Older Adults: Motivations, Challenges and Preferences}

\author{Eugene Tang KangJie}
\affiliation{%
  \institution{National University of Singapore}
  \country{Singapore}
}
\email{eugenetangkangjie@gmail.com}
  
\author{Tianqi Song}
\affiliation{%
  \institution{National University of Singapore}
  \country{Singapore}
}
\email{tianqi_song@u.nus.edu}

\author{Zicheng Zhu}
\affiliation{%
  \institution{National University of Singapore}
  \country{Singapore}
}
\email{zicheng@u.nus.edu}

\author{Jingshu Li}
\affiliation{%
  \institution{National University of Singapore}
  \country{Singapore}
}
\email{jingshu@u.nus.edu}

\author{Yi-Chieh Lee}
\affiliation{%
  \institution{National University of Singapore}
  \country{Singapore}
}
\email{yclee@nus.edu.sg}



\begin{abstract}

\input{section/0Abstract}
\end{abstract}





\maketitle

\input{section/1Introduction}

\input{section/3Methods}

\input{section/4Results}

\input{section/5Discussion}

\input{section/6Limitations}

\input{section/7Conclusion}

\input{section/10Acknowledgements}

\bibliographystyle{ACM-Reference-Format}
\bibliography{reference}

\appendix

\input{section/8Appendix}

\end{document}

%% file: section/0Abstract.tex
As Artificial Intelligence (AI) becomes increasingly integrated into older adults' daily lives, equipping them with the knowledge and skills to understand and use AI is crucial. However, most research on AI literacy education has focused on students and children, leaving a gap in understanding the unique needs of older adults when learning about AI. 
To address this, we surveyed 103 older adults aged 50 and above (Mean = 64, SD = 7). Results revealed that they found it important and were motivated to learn about AI because they wish to harness the benefits and avoid the dangers of AI, seeing it as necessary to cope in the future. However, they expressed learning challenges such as difficulties in understanding and not knowing how to start learning AI. Particularly, a strong preference for hands-on learning was indicated. We discussed design opportunities to support AI literacy education for older adults.

%% file: section/1Introduction.tex
\section{Introduction}

Digital literacy, the ability to use digital technologies to access, evaluate, and share information, has become an essential skill today \cite{reddy2023digital, tinmaz2022systematic}. It empowers individuals to adapt to rapidly changing digital environments, fostering personal growth and social engagement \cite{livingstone2023outcomes, yuan2024digital}. With the growing aging population \cite{world_health_organisation_decade_2023}, digital literacy education for older adults has become a critical topic. Studies have shown that improving digital literacy in older adults not only helps them access vital resources \cite{jurivsic2024enhancing, jung2022health} and protect themselves from online threats \cite{li2024does}, but also increases their independence \cite{karna_multilevel_2022, pihlainen2023older, loe2010doing}, promotes physical and mental well-being \cite{lee2024analysis, choi2013internet}, and helps prevent cognitive decline \cite{tun2010association, quialheiro2022can, klimova2016use}. Recognizing these benefits, governments and social organizations have increasingly prioritized digital literacy education for older adults, contributing to healthy aging and enhancing their overall quality of life.

With the rapid rise of AI technologies and their increasing integration into daily life, AI literacy has become a vital extension of digital literacy \cite{stolpe_artificial_2024, dwivedi2023evolution}. AI literacy encompasses an understanding of AI, the ability to use AI for enhanced value in daily life and the development of critical thinking skills regarding AI \cite{ng_conceptualizing_2021, long_what_2020, yi_establishing_2021}. It could influence older adults in several ways. On the one hand, advancements in AI have driven the development of gerontechnology \cite{aggarwal_novel_2025, caldeira_i_2022, malpani_ageism_2022}, exposing older adults to AI-based technologies more frequently. On the other hand, the loneliness and isolation of older adults could make them more vulnerable to harms related to AI technologies, such as online scams \cite{james2014correlates, xing2020vulnerability, brashier2020aging}. Therefore, it is crucial for older adults to gain the capacity to access and use AI technologies for their needs and to recognize and navigate these potential AI-related dangers.

Despite its importance, educating older adults in AI literacy remains a significant challenge. Reasons include older adults' fear of technology \cite{di2020technophobia, an2024older}, cognitive differences that affect their ability to learn \cite{brito_assessing_2023, yu2020internet}, and a lack of social support during the learning process \cite{tsai_social_2015, kuoppamaki2022enhancing, hunsaker2019he}. However, we acknowledge that older adults are a heterogeneous group \cite{hill2015older} and these factors do not apply universally. Yet, it is still crucial to recognize these reasons because they represent actual difficulties faced by segments of the older adult population.

While recent studies have explored methods to enhance AI literacy education, these efforts have primarily focused on students and children \cite{laupichler_artificial_2022, almatrafi_systematic_2024}. For example, digital story writing has been used in K-12 classrooms to help students understand AI concepts \cite{ng_using_2022}. However, the research specifically targeting older adults remains limited.
Meanwhile, some other initiatives have aimed to support older adults in using AI technologies \cite{rodriguez-martinez_qualitative_2023, desai_ok_2023}, such as incorporating tutorial-based learning phases to teach the use of AI-enabled, speaker-based voice assistants \cite{kim_exploring_2021}. Yet, these studies often focus on specific AI products, restricting their applicability to a broader range of AI technologies.
Additionally, there is a notable gap in addressing older adults' awareness of the potential harms associated with AI, such as misuse or exploitation. Developing comprehensive strategies for AI literacy education that address these gaps is essential to ensuring older adults can confidently and safely engage with AI technologies.

Given the importance of AI literacy for older adults and the existing research gap in this area, we aim to explore how to educate older adults in AI literacy. To address this gap, we began by understanding their motivations (if any) for learning about AI to inform the design of an AI literacy curriculum. Additionally, we investigated their previous learning experiences and preferences to derive actionable insights for implementing an effective AI literacy education process. Based on these objectives, we formulated the following research questions (RQs):

\begin{itemize}
    \item \textbf{RQ1}: What are some possible factors motivating older adults to learn about AI?
    \item \textbf{RQ2}: What are some challenges that older adults face in learning about AI?
    \item \textbf{RQ3}: What are some learning preferences for older adults when it comes to learning about AI?
\end{itemize}

We conducted an online survey study with older adults (N=103). Our findings revealed that older adults generally perceive AI literacy as important and demonstrate a strong motivation to learn it. We identified key motivations, challenges, and preferred learning styles associated with acquiring AI literacy. Based on these findings, we discussed design implications for future AI literacy education. These contributions provide empirical insights for the Human-Computer Interaction (HCI) community, offering guidance for developing AI literacy programs tailored to the unique needs, preferences, and challenges of older adults.

%% file: section/3Methods.tex
\section{Methods}
\subsection{Participants}
In our study, we consider older adults as "individuals aged 50 and above". This definition aligns with organizations such as The American Association of Retired Persons (AARP) \cite{aarp_aarp_2024} and National Seniors Australia \cite{national_seniors_australia_national_2025}. Additionally, the classification of older adults as 50 and above is used in existing literature and research studies \cite{chen2023sport, frechman2022plan, lee2021association, national2020social}.

Prior to recruitment, we sought approval from the Departmental Ethics Review Committee (DERC) at the university. Survey participants were then recruited via email through a local social service organization. The survey was conducted entirely online via Qualtrics. Participants accessed the survey by clicking the survey link provided in the email.

\subsection{Survey Design}
\label{survey-design}
The survey consisted of three sections: (1) Study introduction and consent, (2) main survey, and (3) follow-up demographic information. The introduction and consent section explained that the purpose of the survey was to understand what older adults think of AI literacy education and to learn more about their experiences in gaining information on AI. It also informed participants that they would receive a reimbursement of \$4.50 for completing the survey. 

Next, the main survey addressed the three research questions:
\begin{itemize}
    \item \textbf{RQ1}: To assess participants' motivation to learn about AI, participants rated their motivation levels using a scale adapted from the Motivation-to-Learn Scale \cite{gorges2016likes}. Also, they rated their perceived importance of AI literacy education. We developed statements focused on understanding participants' perceived importance of AI literacy education in general and for specific domains of AI using a modified version of Suh's and Ahn's scale that measures student attitudes toward AI \cite{suh2022development}. We chose three domains relevant to the lives of older adults: Healthcare \cite{barbaccia2022mature, ma2023artificial}, social media \cite{auxier2021social, liu2023ai, nguyen2022deep}, and lifelong learning \cite{kakulla_lifelong_2022, law2023evaluating, fang2024does}. In addition, participants answered an open-ended question about the factors influencing their motivations to learn about AI.
    
    \item \textbf{RQ2}: For challenges in learning AI, participants first responded to a preliminary multiple-choice question to indicate their previous methods for learning about AI. This was followed by an open-ended question asking them to describe the difficulties they encountered when learning or gaining information about AI. Since this question was exploratory, open-ended responses were deemed appropriate to capture a wide range of experiences.
    
    \item \textbf{RQ3}: For learning preferences in studying AI, participants selected one of four options describing how they would prefer to learn about AI, assuming they were to take a class on the subject. Each option corresponded to one of the learning styles described in Kolb’s Learning Theory \cite{kolb2005kolb}: Accommodating, Diverging, Converging, or Assimilating (Table \ref{tab:survey-question-kolb-learning-theory}). Additionally, participants shared further thoughts through an open-ended question on ways to make their learning experiences more enjoyable, imagining if they were to start learning about AI.
\end{itemize}

\begin{table*}[t]
    \centering
    \small
    \renewcommand{\arraystretch}{1.6} 
    \begin{tabular}{>{\columncolor{LightGray}}p{2cm}p{6cm}p{8cm}}
        \hline
         \textbf{Learning Style} & \textbf{Description}    & \textbf{Option Describing the Learning Style}\\
         \hline
         Accommodating & Focuses on a practical hands-on, experiential approach instead of theory.& I want to learn about AI through hands-on activities with AI applications, discovering more about AI as I experiment.\\
        
         Diverging & Prefers to watch people do first, gather information, then use it to solve problems.& I want to learn about AI by observing how others use AI and then reflecting on how I might use it myself.\\
        
        Converging & Learn first, then find solutions to practical issues, focused on problem-solving.      & I want to understand AI concepts in theory first, then explore potential ways that I can use AI before actually using it.\\
        
        Assimilating & Focuses on theory, ideas and concepts, requiring clear explanations rather than hands-on.& I want to attend a structured class that explains AI concepts clearly and theoretically, helping me to understand the foundations of AI.\\
        \hline
    \end{tabular}
    \caption{Survey question that describes learning styles based on Kolb's Learning Theory.}
    \label{tab:survey-question-kolb-learning-theory}
\end{table*}

Finally, the demographic section collected basic information such as age, gender, education level, employment status, recent job, frequency of technology use at work and so on. 
Additionally, we assessed participants' perceptions of AI, digital literacy competency \cite{choi2023everyday, roque2018new, boot2015computer}, readiness to accept technology \cite{parasuraman2015updated}, and perceived self-efficacy in learning about AI \cite{schwarzer1995generalized}. Perceptions of AI were included because this factor is likely to shape participants' motivations to learn about AI. Digital literacy competency, readiness to accept technology and perceived self-efficacy in learning about AI were measured as they could potentially influence the learning preferences of participants when learning about AI. All items in this section, except for the basic information, were assessed using Likert scales adapted from previous studies (see Table \ref{tab:sections-in-survey}). Full survey questions can be found in Appendix \ref{app:full-survey}.

\subsection{Analysis}
\subsubsection{Descriptive Analysis}
For the single-choice questions and multiple-choice questions, we created histograms to determine the count and proportion of participants that selected the corresponding options.
For the Likert scale questions, statements belonging to a single matrix are coded from 1 to 5 (1 - Strongly disagree, 2 - Disagree, 3 - Neutral, 4 - Agree, 5 - Strongly agree). Reverse coding is applied for negatively-phrased statements. Thereafter, we calculated the average score for the statements belonging to the same matrix to arrive at a final score for that matrix.

\subsubsection{Qualitative Analysis}
For the freeform text responses, two researchers analyzed them and performed inductive thematic analysis \cite{braun_using_2006, king_using_2004}. The first researcher created initial codes and grouped relevant content under each code. The second researcher reviewed the codes and discussed with the first researcher to resolve disagreements. The final codes were then organized through affinity diagramming to inductively identify broad themes.

%% file: section/4Results.tex
\section{Results}

\begin{figure*}[t]
    \centering
    \includegraphics[width=0.9\linewidth]{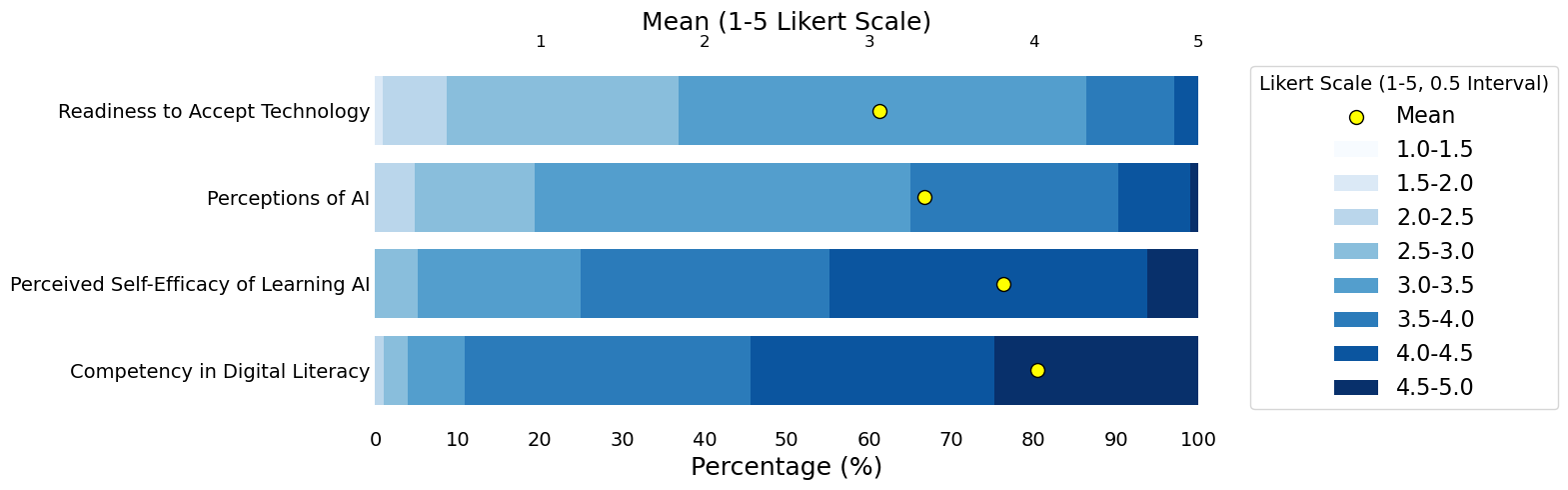}
    \caption{Distribution of scores for Likert-scale survey questions on demographic information. The x-axis (bottom) shows the percentage distribution of participants across Likert-scale intervals, and the x-axis (top) indicates the mean score for each item. Colors represent different score intervals, ranging from light blue (1.0–1.5) to dark blue (4.5–5.0). The yellow circle on each bar marks the mean score for the corresponding item. The mean scores, from top to bottom, are 3.06, 3.34, 3.82, and 4.02.}
    \label{fig:demographic-information}
\end{figure*}

\subsection{Descriptive Results}

We first present descriptive data of the participants' demographic information (see Appendix \ref{app:demographics} for full details). There were a total of 103 (N = 103) participants and they all belonged to the same country. The average survey completion time is 23.66 minutes (SD = 11.84,). 37\% (N = 38) of participants are male, while 63\% (N = 65) are female. The average self-reported age among the participants is 64. The median age is 64, with the 25\textsuperscript{th} and 75\textsuperscript{th} percentiles being 59 and 70 respectively (IQR = 10.5). Highest education level attained ranges from secondary school to Master’s degree. 64\% (N = 66) of participants are currently not working. 66\% (N = 68) of participants indicated that they needed to use technology in their most recent job on a daily basis. 

The participants' digital literacy background is summarized in Figure \ref{fig:demographic-information}. On average, participants demonstrated a digital proficiency score of 4.03 (SD = 0.51, Median = 4) and a readiness to accept technology score of 3.06 (SD = 0.42, Median = 3.06), suggesting they were comfortable with basic digital skills but less prepared to adopt new technologies. Their perceptions of AI yielded an average score of 3.34 (SD = 0.47, Median = 3.42), reflecting attitudes that ranged from slightly negative to neutral. The perceived self-efficacy for learning AI had an average score of 3.81 (SD = 0.57, Median = 3.875), indicating a generally neutral level of confidence in their ability to engage with AI literacy education.

\subsection{Older Adults' Learning Motivations (RQ1)}
\begin{figure*}[t]
    \centering
    \includegraphics[width=\linewidth]{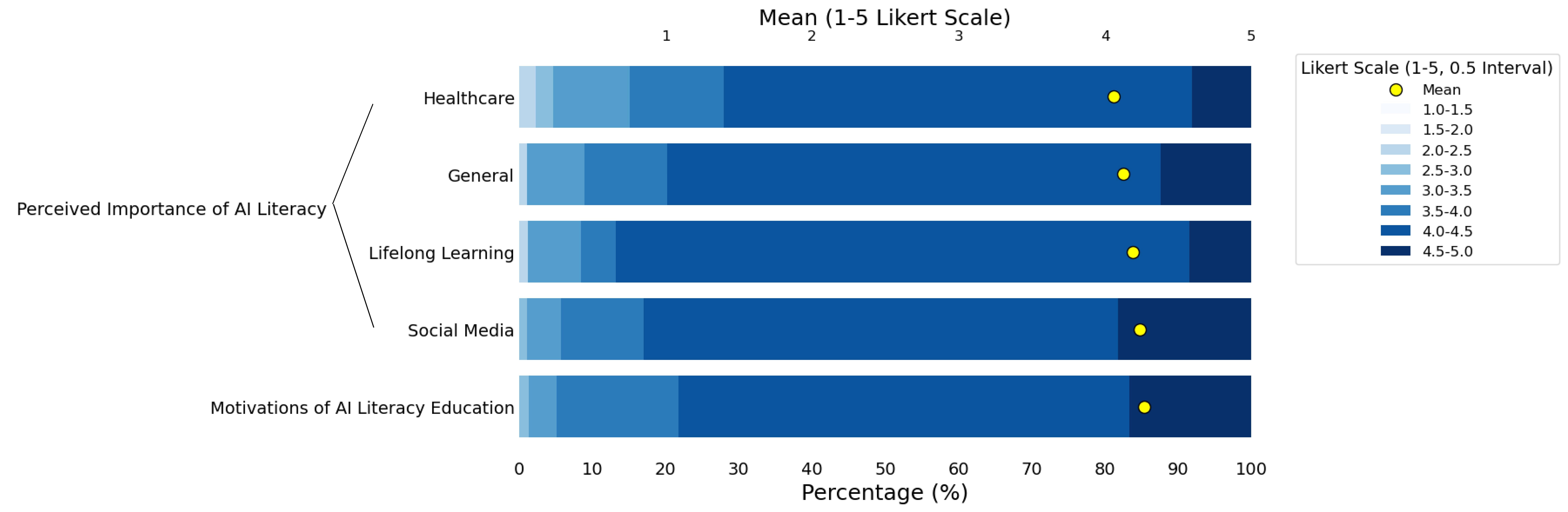}
    \caption{Distribution of scores for Likert-scale survey questions about participants' perceived importance of AI literacy education and motivation to learn AI (RQ1). The x-axis (bottom) shows the percentage distribution of participants across Likert-scale intervals, and the x-axis (top) indicates the mean score for each item. Colors represent different score intervals, ranging from light blue (1.0–1.5) to dark blue (4.5–5.0). The yellow circle on each bar marks the mean score for the corresponding item. The mean scores, from top to bottom, are 4.06, 4.13, 4.19, 4.24 and 4.27.}
    \label{fig:rq1-stacked-bar-chart}
\end{figure*}

In understanding the participants' perceived importance of AI literacy education in general (Fig \ref{fig:rq1-stacked-bar-chart}), the average score is 4.13 (SD = 0.51, Mean = 4), with 82.52\% of participants (N = 85) having a score in [4, 5]. This suggests that the majority of participants find it important to participate in AI literacy education.
We assessed the level of motivation (Fig \ref{fig:rq1-stacked-bar-chart}) and the average score is 4.27 (SD = 0.53, Median = 4). 83.50\% (N = 86) of participants have a score between [4, 5], reflecting \textbf{a high level of motivation to participate in AI literacy education}. The high level of motivation was consistent across age groups (see Appendix \ref{app:motivation-and-age}) and education levels (see Appendix \ref{app:motivation-and-education}). Correlation analysis between the level of motivation and other demographic characteristics can also be found in Appendix \ref{app:motivation-and-other-demographics}. We explored the factors that motivate the participants to pursue AI literacy education (see Table \ref{tab:factors_motivation}).

\begin{table*}
    \centering
    \small
    \renewcommand{\arraystretch}{1.6} 
    \begin{tabular}{>{\columncolor{LightGray}}p{4cm}p{6cm}p{6cm}}

    \hline
    \textbf{Motivation} & \textbf{Description} & \textbf{Sample Quotes} \\
        \hline
        To harness the benefits of AI (N=51, 49.51\%) 
        & {Participants recognize that AI could bring benefits to their lives. They want to learn AI to maximize its benefits. However, they emphasize the importance of real-world applications and to know how AI literacy education could help them in their daily lives.} 
        & {- \textit{“Practical for use is important, instead of theoretical on what it can do.”} (P10)\newline
           - \textit{"I will be very happy if I can apply what I learn in my daily life. For example, I can use Suno AI to create songs/music, I can use AI to create artistic collage of my holiday photos."} (P57)\newline
           - \textit{“Learning AI could help me improve my quality of life in many ways.”} (P59)\newline
           - \textit{“I would like to know how AI could benefit my life, my loved ones, friends, society, and the country in general, being mindful also of the pitfalls of AI.”} (P72)
           }\\
        \hline

         To cope in the future (N=42, 40.78\%)
        & {Participants believe AI will become an integral part of daily life. They fear being left behind in society and want to learn AI to maintain independence and stay connected.} 
        & {- \textit{“It is important as in the future we will be using AI in our daily life activities.”} (P27)\newline
           - \textit{“It is important as new technology advances, and if you do not keep up the pace, you will find yourself outdated and hopeless.”} (P29)\newline
           - \textit{“It is a new way of life eventually, you cannot avoid it, and trying to ignore this new trend is going to be a big mistake.”} (P77)
            }\\
        \hline
        To avoid the dangers of AI (N=29, 28.16\%)
        & {Participants want to learn about AI because they recognize that AI brings new dangers to their lives. Knowing more about AI could help them avoid such dangers.} 
        & {- \textit{“To avoid pitfalls or scams with awareness of what AI can be used adversely.”} (P23)\newline
           - \textit{“AI is now upcoming and I must learn how to harness it and not be scammed.”} (P32)\newline
           - \textit{"A balance view of how it can be better leveraged as a tool yet being mindful of the threats when it may be abused."} (P95)
           }\\
           \hline
    \end{tabular}
\caption{Factors motivating participants to learn about AI.}
\label{tab:factors_motivation}
\end{table*}

\subsection{Older Adults' Learning Challenges (RQ2)}
The survey first investigated how participants typically obtained information or learned about AI. A significant majority (92.2\%) reported using methods such as search engines, videos, online news, or social networks, while only 7.8\% indicated they had never used any of these sources (see detailed results in Appendix \ref{app:methods-of-learning-about-ai}).
To address the research question (RQ), the survey included an open-ended question inviting participants to describe any challenges they faced in learning about AI. While some participants had limited engagement with learning about AI and thus did not report any challenges, 66.99\% (N = 69) participants identified specific learning challenges that they faced (see Table \ref{tab:learning_challenges}).

\begin{table*}
    \centering
    \small
    \renewcommand{\arraystretch}{1.6} 
    \begin{tabular}{>{\columncolor{LightGray}}p{4cm}p{6cm}p{6cm}}
        \hline
        \textbf{Challenge} & \textbf{Description} & \textbf{Sample Quotes} \\
            \hline
            Difficulties in understanding (N=22, 21.36\%)
            & {Participants expressed the need for someone to explain concepts or instructions to them in a guided manner, as self-guided learning often proved insufficient. Participants also noted that they require additional time to grasp the information provided, especially when it comes to technical aspects of AI which they found difficult to understand.}  
            & {- {\textit{“I am clueless when I touched on the technical aspect of AI. e.g Machine Learning and Data Sciences.”} (P5)\newline
               - \textit{“Challenges may be we do not understand some terminology used that we have not come across before.”} (P35)\newline
               - \textit{“[I] do not understand by my own.”} (P82)}
            } \\
            \hline
             Not knowing how to start learning AI (N=12, 11.65\%)
            & {Participants expressed that they struggled to find suitable learning avenues and resources. Additionally, some other participants revealed that they were unsure of how to begin learning AI altogether.}  
            & {- {\textit{“I am not sure how to learn AI, it seems to be technology that is so advanced that seniors like myself will find it hard to learn.”} (P7)\newline
               - \textit{“[Not] sure which is an appropriate class to attend.”} (P40)\newline
               - \textit{“Yet to find any platform to learn more about Al right now.”} (P52)\newline
               - \textit{“Do not know what and where to find the information.” (P60)}}
            }\\
            \hline
            Information retrieval (N=11, 10.68\%)
            & {There is a need to rely on multiple sources of information when searching for information to learn about AI, thus requiring participants to navigate the decision-making process of determining which sources are trustworthy and relevant to their specific questions.}  
            & {- {\textit{“Not clear info from internet for self help. Need to keep searching and find the suitable info after trial and error.”} (P41)\newline
               - \textit{“Too much information available. I need to sieve through the information to select the most accurate and useful details needed.”} (P96)}
            } \\
            \hline
            Lack of opportunities for hands-on practice (N=10, 9.71\%)
            & {Participants noted difficulties in finding opportunities to apply or reinforce AI-related knowledge.}  
            & {- {\textit{“After learning, would I be able to practice?”} (P91)\newline
               - \textit{“Lack of follow-up guidance.”} (P98)}
            }\\
            \hline
    \end{tabular}
\caption{Challenges faced by participants when learning about AI.}
\label{tab:learning_challenges}
\end{table*}

\subsection{Older Adults' Learning Preferences (RQ3)}
\label{sec:learning-preferences-and-challenges}

\begin{figure*}
    \centering
    \includegraphics[width=0.8\linewidth]{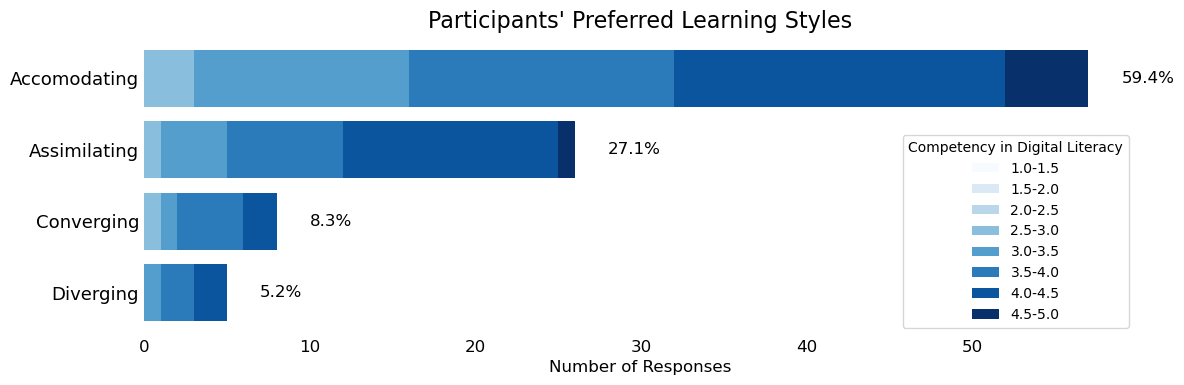}
    \caption{Preferred learning style indicated by participants among options modeled using Kolb's Learning Theory. Colors represent different score intervals of participants' competency in digital literacy, ranging from light blue (1.0–1.5) to dark blue (4.5–5.0). }
    \label{fig:piechart-preferred-learning-style}
\end{figure*}

The survey results reveal a strong preference for the Accommodating learning style, with 59.22\% (N = 61) participants choosing it (see Fig \ref{fig:piechart-preferred-learning-style}). The Accommodating learning style emphasizes hands-on, experiential learning \cite{kolb2005kolb} where participants directly interact with AI applications, learning through experimentation instead of heavily focusing on theoretical AI concepts. Beyond learning styles, participants were asked about aspects that would make their learning experiences enjoyable and helpful when starting to learn about AI (see Table \ref{tab:learning_preferences}).

\begin{table*}
    \centering
    \small
    \renewcommand{\arraystretch}{1.6} 
    \begin{tabular}{>{\columncolor{LightGray}}p{4cm}p{6cm}p{6cm}}
        \hline
        \textbf{Preference} & \textbf{Description} & \textbf{Sample Quotes} \\
            \hline
            Hands-on practice (N=41, 39.81\%)
            & {An overwhelming aspect mentioned was the need to have hands-on practice. This preference closely aligns with the preferred Accommodating learning style highlighted, further emphasizing participants’ strong inclination to learn by doing.}  
            & { - \textit{“I like hands-on learning, practice helps me gain confidence in AI.”} (P61)\newline
                - \textit{“To do hands on, and have more tutorials, ways that application is used.”} (P102)\newline
                - \textit{"Hands on and seeing it in application myself."} (P103)
            } \\
            \hline
             Quality social support (N=20, 19.42\%)
            & {Participants desired for quality social support during the learning process. There are 2 parts. The first was the importance of having patient and friendly teachers to guide them in learning AI. The second part was about wanting to learn alongside and interact with fellow learners. Participants valued opportunities for collaboration and shared learning.}  
            & {- {\textit{“Small group like 5-10 pax will be advisable so that sharing can be encouraged and explored.”} (P15) \newline
               - \textit{“A friendly and knowledgeable coach”} (P31) \newline
               - \textit{“Interaction with the instructor/other participants and sharing of knowledge”} (P83)}
            }\\
            \hline
            Illustrations and examples (N=11, 10.68\%)
            & {Participants expressed a preference for illustrations and examples when learning about AI. These often involved demonstrating how AI could be useful to them. The concept of usefulness was closely tied to examples that connected AI to participants’ personal lives or areas of interest.}  
            & {- {\textit{“[Illustrations] and examples on how AI can be used to offer solutions to some issues or situations, whether it is at work or at home.”} (P21)\newline
               - \textit{"[Being] able to see the immediate benefits of AI during the lesson if the instructor shows some examples”. (P59)}\newline
               - \textit{“Real life and useful examples.”} (P76)} 
            }\\
            \hline
    \end{tabular}

\caption{Learning preferences of participants for AI literacy education.}
\label{tab:learning_preferences}
\end{table*}

%% file: section/5Discussion.tex
\section{Discussion}

\subsection{AI Literacy Education in an Aging Age}
Our research contributes to the broader field of research regarding older adults and digital literacy education. Much of the existing work on digital literacy focus on basic digital usage such as teaching older adults how to perform smartphone tasks \cite{jin2024exploring, gao2024easyask, leung2012older, liu2019help} or using a computer \cite{gonzalez2015ict, mayhorn2004older}. However, the technology landscape has evolved significantly and many of these tasks have become mainstream \cite{pang2021technology}. Older adults are now more exposed to digital technology, especially as it is integrated into essential aspects of daily life, such as in digital payments \cite{ramayanti2024exploring} and government services \cite{gil2018digital}. In contrast, AI literacy education raises different considerations. AI is not yet as indispensable as everyday technologies like smartphones and the Internet. It is crucial to explore whether older adults are interested in or find it important to learn about AI where these attitudes and perceptions are fundamental to AI literacy education for older adults. 

Our findings highlight that older adults value AI literacy education and are motivated to learn, underscoring the need for continued and focused efforts by researchers and society to promote AI literacy. 
Previous studies have noted that older adults may face challenges with advanced digital literacy education due to limited prior experience with technology, and suggest to lower expectation on what digital literacy to teach older adults \cite{vercruyssen2023basic}. While we agree on the importance of designing educational approaches that align with their existing skills, we emphasize the need to empower older adults to engage with advanced technologies like AI. Such efforts can enable them to fully harness the benefits of technology, prepare for future challenges, and mitigate potential risks.

\subsection{Design Opportunities for Future Work}
\subsubsection{Teaching through Illustrating AI's Relevance}
One key reason participants find it important to learn about AI is their recognition of the potential benefits and dangers AI could bring to their lives. This is reflected in their preference for illustrations and examples that can help them recognize the practical value of AI. 
Our finding aligns with Bhat et al., who found that adults without technical backgrounds, especially those in the workforce, preferred learning experiences tied to real-world applications \cite{bhat2024designing}. For older adults, particularly those no longer working, AI literacy education should adopt an application-driven approach, demonstrating AI’s relevance through everyday scenarios.

This design opportunity of teaching through illustrating AI's relevance can be incorporated into AI literacy education tools designed for older adults. An example could be related to one of the challenges identified from our survey results - The difficulty older adults face in retrieving AI-related information. 
While existing tools powered by large language models (LLMs) like ChatGPT can serve as resource locators, their explanations tend to be theoretical and lengthy, limiting their effectiveness for older adults. To address this, persona-based AI assistants and tailored prompts could present information in familiar, real-world contexts \cite{ebner2023financial, herrera2024bridging}. Structuring conversations around real-life scenarios and emphasizing practical benefits—such as recognizing scams—could enhance both comprehension and motivation to learn.

\subsubsection{Fostering Hands-on Learning}
The majority of participants indicated a preference for the Accommodating learning style, favoring exploratory learning through hands-on activities with AI applications. The preference is further supported by one of the main learning challenges identified: The lack of opportunities for hands-on practice. This reveals a design opportunity in AI literacy education for older adults where there can be exploration regarding how older adults can learn via doing.

A direct approach to hands-on learning is guiding older adults to use existing AI applications that are relevant to their daily lives. For example, older adults still in the workforce could benefit from experimenting with AI tools like Copilot in Microsoft applications. However, AI literacy extends beyond knowing how to use AI tools, where it also involves understanding AI concepts and critically evaluating its impact \cite{ng_conceptualizing_2021, long_what_2020, yi_establishing_2021}. 
Teaching tool functionality alone does not address these deeper aspects, highlighting the need for interactive learning experiences that integrate both practical engagement and conceptual understanding.

\subsubsection{Leveraging Social Learning}
Previous research highlights that older adults utilize social learning, receiving assistance from their social networks when learning new technologies \cite{pang2021technology, sharifi2023senior, mendel2021exploratory}. The emphasis on the social aspect is reflected in our survey results. Participants expressed a preference for learning from patient and approachable teachers and engaging with fellow learners for peer learning.

In terms of program design, these insights point to the potential of exploring a peer learning paradigm for AI literacy education tools. 
Peer learning combines both teaching and learning, fostering a collaborative learning environment \cite{chen2020teaching}. While peer learning with non-human entities, such as chatbots, has been explored in academic literature \cite{jin2024teach, lee2021curiosity, looi2008design}, research on its application for older adults remains limited.

It is worth investigating how non-human entities like chatbots and robots could be designed to incorporate a peer learning approach to support AI literacy education for older adults. 
A chatbot, for example, could adopt the persona of a fellow older adult learner—sharing perspectives and experiences while also learning from the user. This approach could be especially valuable for older adults who lack access to structured classroom settings, providing a socially engaging experience tailored to their needs.

%% file: section/6Limitations.tex
\section{Limitations}

Our study has several limitations. Firstly, participants were recruited from a single social service organization, which may introduce selection bias due to the lack of diversity in the social and economic backgrounds of the participants. This could limit the generalizability of our results. Secondly, the survey did not assess participants' baseline understanding of AI. Thus, their responses were based on their personal interpretations of what AI is, which could be influenced by the sources they used to learn about AI (Appendix \ref{app:methods-of-learning-about-ai}). This introduces the possibility of variation in how participants perceived AI while completing the survey. Thirdly, when eliciting older adults' preferred learning style when it comes to learning about AI, the survey question was framed under the assumption that participants would take a class on the subject (Section~\ref{survey-design}). This may have implied a formal, structured learning environment and therefore, could have overlooked informal learning settings where science and technology education often takes place \cite{falk2011lifelong}.

Given these limitations, we encourage future research to explore older adults' learning preferences and challenges in both formal and informal learning settings. Additionally, future studies should involve older adults from a more diverse range of backgrounds to ensure a broader representation of the older adult population.

%% file: section/7Conclusion.tex
\section{Conclusion}
This paper presents the findings of an online survey conducted among 103 older adults aged 50 and above on AI literacy education. The results show that participants find it important to learn about AI and are motivated to do so. They want to harness AI’s benefits, but the benefits and the applicability to their daily lives ought to be clearly illustrated. Key learning challenges were discussed, including difficulties with understanding and information retrieval which highlight opportunities for designing more accessible and supportive learning environments. Our findings emphasize that AI literacy education for older adults is a timely and significant topic worthy of future exploration. It serves as a starting point for future research in this topic and we encourage AI literacy educators to investigate potentially effective pedagogies tailored to older adults.

%% file: section/10Acknowledgements.tex
\begin{acks}
    This research was supported by Yale-NUS seed grant (A-8001353-00-00).
    We sincerely thank Family Central, Fei Yue Community Services, for their support in our work and their assistance in recruiting survey participants.
\end{acks}

%% file: section/8Appendix.tex
\section{Appendix}

\subsection{Full Survey Questions}
\label{app:full-survey}
\begin{enumerate}

\begin{table*}[ht]
    \centering
    \small
    \renewcommand{\arraystretch}{1.6} 
    \begin{tabular}{p{3cm} p{5cm} p{2cm} p{5cm}}
    \hline
    \toprule
     \textbf{Focus} & \textbf{Subsections} & \textbf{ Questions} & \textbf{Scales and Theories Adapted} \\
     \hline
    \textbf{RQ1:} Motivation to learn about AI & Perceived importance of AI literacy education & Q1 - Q4 & Scale measuring student attitudes toward AI 
    \cite{suh2022development} \\
    & Open-ended Question & Q5 & $-$ \\
           & Motivations of AI literacy education & Q6 & Motivation-to-Learn scale \cite{gorges2016likes}\\
    & Open-ended Question & Q7 & $-$ \\
    \hline
    \textbf{RQ2:} Challenges in learning about AI & Previous methods participants used to learn about AI & Q8 & $-$ \\ 
    & Open-ended Question & Q9 & $-$ \\
    \hline
    \textbf{RQ3:} Learning preferences to learn about AI & Learning about AI & Q10 & Kolb's Learning Theory \cite{kolb2005kolb}\\
    & Open-ended Question & Q11 & $-$ \\
    \hline
    \textbf{Demographics:} Basic Demographics and Digital Literacy Background & 
Basic Demographics & Q12 - Q18 & $-$ \\
    & Competency in digital literacy & Q19 - Q20 & Everyday digital literacy questionnaire for older adults \cite{choi2023everyday}, Mobile device proficiency questionnaire \cite{roque2018new}, Computer proficiency questionnaire \cite{boot2015computer}\\
        & Readiness to accept technology & Q21 & Technology Readiness Index (TRI) 2.0 \cite{parasuraman2015updated}\\
        & Perceptions of AI & Q22 & General Attitudes towards Artificial Intelligence Scale (GAAIS) \cite{schepman2020initial}.\\
        & Perceived self-efficacy in learning about AI & Q23 & General self-efficacy scale \cite{schwarzer1995generalized}\\
    \hline
    \end{tabular}
    \caption{Structure of the survey design, incorporating references from established scales and theories.}
    \label{tab:sections-in-survey}
\end{table*}

 \item \textbf{Perceived Importance of AI Literacy Education}
    \begin{itemize}
        \item \textbf{Q1:} To what extent do you agree with each of the following statements? For each statement, you may select one of the following: Strongly disagree, Disagree, Neutral, Agree, Strongly agree.
        \begin{itemize}
            \item I find that AI is something worth learning.
            \item Classes on AI are important.
            \item I think seniors should learn about AI.
            \item I will need AI in my life in the future.
        \end{itemize}

        \item \textbf{Q2:} To what extent do you agree with each of the following statements? For each statement, you may select one of the following: Strongly disagree, Disagree, Neutral, Agree, Strongly agree.
        \begin{itemize}
            \item I find that it is important to learn how AI can support me in making healthcare decisions.
            \item I care about knowing how AI can help me to live independently.
            \item I care about developments in AI that can help me manage age-related medical problems, such as diabetes or hypertension.
        \end{itemize}

         \item \textbf{Q3:} To what extent do you agree with each of the following statements? For each statement, you may select one of the following: Strongly disagree, Disagree, Neutral, Agree, Strongly agree.
        \begin{itemize}
            \item I care about the dangers AI brings to social media users.
            \item It is important for me to learn how to use AI tools to avoid scams on social media.
            \item I find it important to be exposed to AI tools that can make my social media experience more enjoyable.
        \end{itemize}

        \item \textbf{Q4:} To what extent do you agree with each of the following statements? For each statement, you may select one of the following: Strongly disagree, Disagree, Neutral, Agree, Strongly agree.
        \begin{itemize}
            \item I care about knowing how AI can solve the problems that I face when learning about new things.
            \item I find that it is important to learn how to use AI to make it easier for me to learn new things.
            \item It is important for me to know how AI can make it easier for me to access information that I need in my daily life.
        \end{itemize}

         \item \textbf{Q5:} Do you find that it is important for you to learn about AI? Why or why not? Please share your thoughts in at least 1 - 2 sentences.
    \end{itemize}

 \item \textbf{Motivations of AI Literacy Education}
    \begin{itemize}
        \item \textbf{Q6:} To what extent do you agree with each of the following statements? For each statement, you may select one of the following: Strongly disagree, Disagree, Neutral, Agree, Strongly agree.
        \begin{itemize}
            \item I want to learn about AI.
            \item If I do not understand something about AI, I am willing to look for information to make it clearer.
            \item I will participate in a class that teaches about AI.
            \item I am interested in the future developments of AI.
        \end{itemize}


        \item \textbf{Q7:} If you are deciding whether to learn about AI, what are some factors that will influence your decision? Please share your thoughts in at least 1 - 2 sentences.
    \end{itemize}

 \item \textbf{Learning about AI}
    \begin{itemize}
        \item \textbf{Q8:} Which of the following ways have you used to learn about or gain information on AI? (Select all that apply)
        \begin{itemize}
            \item[$\square$] Physical classes (e.g. At community centers or libraries)
            \item[$\square$] Online courses (e.g., Coursera)
            \item[$\square$] Videos (e.g., YouTube)
            \item[$\square$] Social networks (e.g, Learn about AI from family members and friends)
            \item[$\square$] Newspapers or online news (e.g., Reading news about AI)
            \item[$\square$] Search engines (e.g., Search on Google to learn about AI)
            \item[$\square$] Social media (e.g., View AI-related content on Facebook or Instagram)
            \item[$\square$] I do not use any of the above
            \item[$\square$] Others: \_\_\_\_
        \end{itemize}

        \item \textbf{Q9:} Did you face any challenges when learning about or gaining information on AI? If so, what were some challenges you faced? Please share your thoughts in at least 1 - 2 sentences.

        \item \textbf{Q10:} Imagine you are going to participate in a class that teaches about AI. Choose the statement that you agree with the most.
        \begin{itemize}
            \item[$\circ$] I want to learn about AI through hands-on activities with AI applications, discovering more about AI as I experiment.
            \item[$\circ$] I want to learn about AI by observing how others use AI and then reflecting on how I might use it myself.
            \item[$\circ$] I want to understand AI concepts in theory first, then explore potential ways that I can use AI before actually using it.
            \item[$\circ$] I want to attend a structured class that explains AI concepts clearly and theoretically, helping me to understand the foundations of AI.
        \end{itemize}
        
        \item\textbf{Q11:} Imagine you want to start learning about AI. What would make the learning experience enjoyable or helpful for you? You can share anything that comes to your mind, including the learning format (e.g. Online self-directed learning or in-person classes), the learning style (e.g. Hands-on learning) or anything else you would want to see. Please share your thoughts in at least 1 - 2 sentences.
    \end{itemize}

\item \textbf{Basic Demographics}
    \begin{itemize}
        \item \textbf{Q12:} What is your age?  
        \item \textbf{Q13:} How do you describe your gender identity?  
        \begin{itemize}
            \item[$\circ$] Male  
            \item[$\circ$] Female
            \item[$\circ$] Others
            \item[$\circ$] Prefer not to say  
        \end{itemize}
        \item \textbf{Q14:} What is the highest level of education you have completed?
        \begin{itemize}
            \item[$\circ$] Primary School
            \item[$\circ$] Secondary School
            \item[$\circ$] Tertiary Education
            \item[$\circ$] Bachelor's Degree
            \item[$\circ$] Master's Degree
            \item[$\circ$] Ph.D. or Higher
            \item[$\circ$] Prefer not to say
        \end{itemize}
        \item \textbf{Q15:} Are you currently employed?
        \begin{itemize}
            \item[$\circ$] Yes
            \item[$\circ$] No
        \end{itemize}
        \item \textbf{Q16:} What is your most recent job?
        \item \textbf{Q17:} How often did your most recent job require you to use technology?
        \begin{itemize}
            \item[$\circ$] Daily
            \item[$\circ$] Occasionally
            \item[$\circ$] Rarely
            \item[$\circ$] Not at all
        \end{itemize}
        \item \textbf{Q18:} Could you briefly describe how you used technology in your most recent job?
    \end{itemize}

\item \textbf{Competency in Digital Literacy}
    \begin{itemize}
        \item \textbf{Q19:} What digital devices do you use? (Select all that apply)
        \begin{itemize}
            \item[$\square$] Smart phone
            \item[$\square$] Tablet
            \item[$\square$] Computer (Desktop or Laptop)
            \item[$\square$] Smart Watch
            \item[$\square$] Smart TV
            \item[$\square$] Others: \_\_\_\_
        \end{itemize}

        \item \textbf{Q20:} To what extent do you agree with each of the following statements? For each statement, you may select one of the following: Strongly disagree, Disagree, Neutral, Agree, Strongly agree.
        \begin{itemize}
            \item I am able to find information that I need on the Internet.
            \item I can communicate with others through the Internet (e.g, Emails, messaging apps, social media).
            \item I know where to get help or how to ask for help when I face problems using technology.
            \item I can create a document using digital devices (e.g, Word document).
            \item I know how to judge whether information from the Internet is reliable or not.
            \item I am able to save Internet documents, photos, or video files that I find.
            \item I know how to delete files stored on my digital devices.
            \item I can independently troubleshoot issues related to device/app operation.
        \end{itemize}
    \end{itemize}

    \item \textbf{Readiness to Accept Technology}
    \begin{itemize}
        \item \textbf{Q21:} To what extent do you agree with each of the following statements? For each statement, you may select one of the following: Strongly disagree, Disagree, Neutral, Agree, Strongly agree.
        \begin{itemize}
            \item New technologies contribute to a better quality of life.
            \item Technology gives me more freedom of mobility.
            \item Technology gives people more control over their daily lives.
            \item Technology makes me more productive in my personal life.
            \item Other people come to me for advice on new technologies.
            \item In general, I am among the first in my circle of friends to acquire new technology when it appears.
            \item I can usually figure out new high-tech products and services without help from others.
            \item I keep up with the latest technological developments in my areas of interest.
            \item When I get technical support from a provider of a high-tech product or service, I sometimes feel as if I am being taken advantage of by someone who knows more than I do.
            \item Technical support lines are not helpful because they don’t explain things in terms I understand
            \item Sometimes, I think that technology systems are not designed for use by ordinary people.
            \item There is no such thing as a manual for a high-tech product or service that’s written in plain language.
            \item People are too dependent on technology to do things for them.
            \item Too much technology distracts people to a point that is harmful.
            \item Technology lowers the quality of relationships by reducing personal interaction.
            \item I do not feel confident doing business with a place that can only be reached online.    
        \end{itemize}
    \end{itemize}

    \item \textbf{Perceptions of AI}
    \begin{itemize}
        \item \textbf{Q22:} To what extent do you agree with each of the following statements? For each statement, you may select one of the following: Strongly disagree, Disagree, Neutral, Agree, Strongly agree.
        \begin{itemize}
            \item I am interested in using artificially intelligent systems in my daily life.
            \item There are many beneficial applications of Artificial Intelligence.
            \item Artificial Intelligence can provide new economic opportunities for my country.
            \item I am impressed by what Artificial Intelligence can do.
            \item Artificially intelligent systems can help people feel happier.
            \item Much of society will benefit from a future full of Artificial Intelligence.
            \item I think Artificial Intelligence is dangerous.
            \item Artificial Intelligence is used to spy on people.
            \item I shiver with discomfort when I think about future uses of Artificial Intelligence.
            \item Artificial Intelligence might take control of people.
            \item I think artificially intelligent systems make many errors.
            \item People like me will suffer if Artificial Intelligence is used more and more.
        \end{itemize}
    \end{itemize}

     \item \textbf{Perceived Self-efficacy in Learning about AI}
    \begin{itemize}
        \item \textbf{Q23:} Suppose you are learning about AI. It could be about understanding what AI is, learning how to use a new AI product or getting to know about the benefits and dangers of AI. To what extent do you agree with each of the following statements? For each statement, you may select one of the following: Strongly disagree, Disagree, Neutral, Agree, Strongly agree.
        \begin{itemize}
            \item If I try hard enough, I can solve difficult problems that arise during the learning process.
            \item It is easy for me to stick to my learning objectives regarding AI and accomplish them.
            \item I am confident that I could deal efficiently with unexpected events when I am learning about AI.
            \item Thanks to my resourcefulness, I know how to handle unforeseen situations while learning about AI.
            \item I can remain calm when facing learning difficulties because I can rely on my coping abilities.
            \item When I am confronted with a problem while learning about AI, I can find several solutions.
            \item If I am stuck with something during the learning process, I can think of something to do.
            \item No matter what comes my way during learning, I am usually able to handle it.
        \end{itemize}
    \end{itemize}

\end{enumerate}

\subsection{Demographics of Survey Participants}
\label{app:demographics}
Table \ref{tab:basic_demographics} provides an overview of the basic demographic characteristics of the survey participants. Table \ref{tab:additional_demographics} presents the number of participants who scored within each bin for the demographic characteristics on digital literacy competency, readiness to accept technology, perceptions of AI and perceived self-efficacy in learning about AI.

\begin{table*}
    \centering
    \small
    \renewcommand{\arraystretch}{1.6} 
    \begin{tabular}{p{6cm}p{4cm}p{4cm}p{2cm}}

    \hline
    \toprule
    \textbf{Demographic} & \textbf{Values} & \textbf{Number of Participants} (N = 103) & \textbf{Percentage} (\%) \\
    \hline

    Gender & Male & 38  & 36.89\\
           & Female & 65 & 63.10\\
    \midrule
    Age & 50 - 54 & 10  & 9.71\\
        & 55 - 59 & 16  & 15.53\\
        & 60 - 64 & 26  & 25.24\\
        & 65 - 69 & 24 & 23.30\\
        & 70 - 74 & 19  & 18.45\\
        & 75 - 80 & 8  & 7.77\\
    \midrule
    Education & Master's degree & 14 & 13.59\\
              & Bachelor's degree & 42 & 40.78\\
              & Tertiary education & 27 & 26.21\\
              & Secondary school & 18 & 17.48\\
              & Primary school & 0 & 0.00\\
              & Prefer not to say & 2 & 1.94\\
    \midrule
    Employment Status & No & 66 & 64.08\\
                      & Yes & 37 & 35.92\\
    \midrule
    Frequency of Technology Use in Most Recent Job & Daily & 68 & 66.02\\
        & Occasionally & 15 & 14.56\\
        & Rarely & 9 & 8.74\\
        & Not at all & 11 & 10.68\\
    \midrule
    \end{tabular}
     \caption{Basic demographic characteristics of survey participants.}
      \label{tab:basic_demographics}
\end{table*}

\begin{table*}
    \centering
    \small
    \renewcommand{\arraystretch}{1.6} 
    \begin{tabular}{p{3.5cm}p{1cm}p{1.5cm}p{1.5cm}p{1.5cm}p{1.5cm}p{1.5cm}p{1.5cm}p{1.5cm}}

    \hline
    \toprule
    \textbf{Demographic}
    & \textbf{[1, 1.5]}
    & \textbf{(1.5, 2]}
    & \textbf{(2, 2.5]}
    & \textbf{(2.5, 3]}
    & \textbf{(3, 3.5]}
    & \textbf{(3.5, 4]}
    & \textbf{(4, 4.5]}
    & \textbf{(4.5, 5]}
    \\
    \hline
   
    Competency in Digital Literacy\newline(Mean = 4.03, SD = 0.51, Median = 4)
    & 0 (0.00\%)
    & 0 (0.00\%)
    & 1 (0.97\%)
    & 3 (2.91\%)
    & 12 (11.65\%)
    & 43 (41.75\%)
    & 28 (27.18\%)
    & 16 (15.53\%)
    \\
    \hline

     Readiness to Accept Technology\newline(Mean = 3.06, SD = 0.42, Median = 3.06)
     & 0 (0.00\%)
     & 1 (0.97\%)
     & 10 (9.71\%)
     & 39 (37.86\%)
     & 39 (37.86\%)
     & 13 (12.62\%)
     & 1 (0.97\%)
     & 0 (0.00\%)
     \\
     \hline

    Perceptions of AI\newline(Mean = 3.34, SD = 0.47, Median = 3.42)
    & 0 (0.00\%)
    & 1 (0.97\%)
    & 5 (4.85\%)
    & 19 (18.45\%)
    & 50 (48.54\%)
    & 22 (21.36\%)
    & 6 (5.83\%)
    & 0 (0.00\%)
    \\
    \hline

    Perceived Self-efficacy in Learning AI\newline(Mean = 3.81, SD = 0.57, Median = 3.875)
    & 0 (0.00\%)
    & 0 (0.00\%)
    & 0 (0.00\%)
    & 13 (12.62\%)
    & 17 (16.50\%)
    & 51 (49.51\%)
    & 11 (10.68\%)
    & 11 (10.68\%)
    \\
    \hline
    \end{tabular}
     \caption{The number of participants that scored within each bin for the demographic characteristics on digital literacy competency, readiness to accept technology, perceptions of AI and perceived self-efficacy in learning about AI. The percentages are calculated as a proportion of the total number of participants (N = 103).}
      \label{tab:additional_demographics}
\end{table*}

\subsection{Additional Survey Findings}

\subsubsection{Methods of Learning About AI}
\label{app:methods-of-learning-about-ai}
\begin{figure*}
    \centering
    \includegraphics[width=0.8\linewidth]{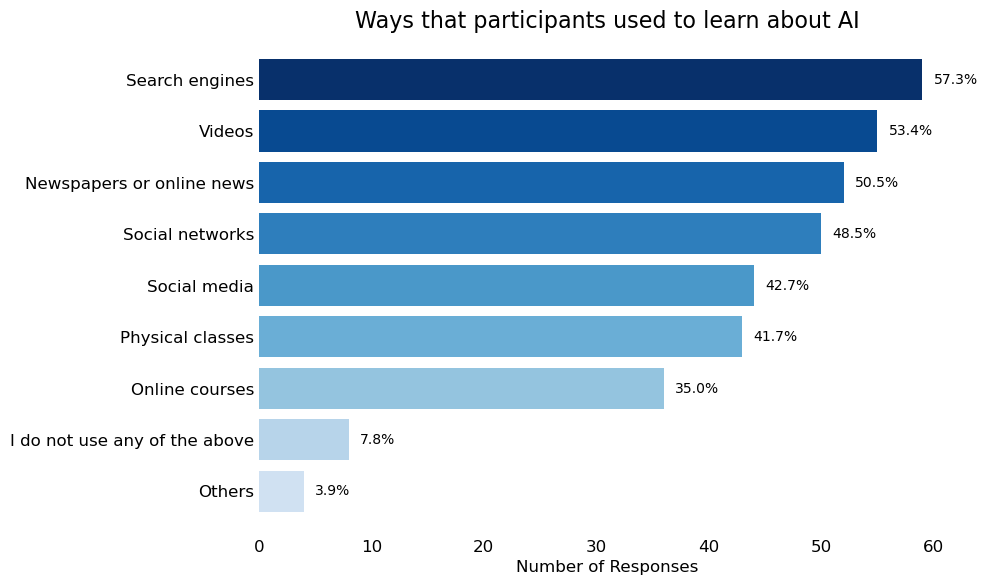}
    \caption{Ways that participants used to gain information or learn about AI.}
    \label{fig:ways-of-learning-ai-bar-chart}
\end{figure*}

This section provides additional insights regarding the methods participants previously used for learning about AI. It does not directly address our research questions but it is related to the learning experiences of older adults. The results (Fig \ref{fig:ways-of-learning-ai-bar-chart}) revealed that the most common approaches included using search engines to explore AI-related topics (N = 59, 57.3\%), watching videos about AI (N = 55, 53.4\%), reading newspapers or online news (N = 52, 50.5\%), and engaging with social networks or social media (N = 50, 48.5\%).

\subsubsection{Motivation and Age}
\label{app:motivation-and-age}
Given that participants range in age from 50 to 80, we explored how the motivation of AI literacy education varies across different age groups. Table \ref{tab:motivations_across_age} provides details of the motivation scores categorized into five-year age groups. While there are some variations, it is noteworthy that the median motivation score across all age groups is 4.00 or higher. Except for the age group of 55 - 59, all other age groups have a 25th percentile of 4.00 and above. This suggests that the motivation to engage in AI literacy education remains consistently high across all age groups and is not concentrated in any one group.

Furthermore, we used Data Analysis ToolPak in Excel to perform regression analysis examining the relationship between participants' age and their motivation scores. The correlation coefficient was 0.12, indicating a weak or negligible correlation. Additionally, the p-value associated with the F statistic is 0.23 which is greater than 0.05, suggesting that age does not have a statistically significant relationship with the motivation score.

\begin{table*}
    \centering
    \small
    \renewcommand{\arraystretch}{1.6} 
    \begin{tabularx}{\textwidth}{p{3cm}p{2cm}p{2cm}p{2cm}p{2cm}p{2cm}p{2cm}}
    \hline
    \toprule
    \textbf{Age Group} & \textbf{Min} & \textbf{Max} & \textbf{25th Percentile} & \textbf{Median} & \textbf{75th Percentile} & \textbf{IQR} \\
    \hline
    50 - 54 (N=10, 9.71\%) & 3.50 & 5.00 & 4.00 & 4.00 & 4.1875 & 0.1875 \\
    55 - 59 (N=16, 15.53\%) & 3.25 & 5.00 & 3.75 & 4.00 & 4.375 & 0.625 \\
    60 - 64 (N=26, 25.24\%) & 3.25 & 5.00 & 4.00 & 4.375 & 5.00 & 1.00 \\
    65 - 69 (N=24, 23.30\%) & 2.75 & 5.00 & 4.00 & 4.00 & 4.75 & 0.75 \\
    70 - 74 (N=19, 18.45\%) & 4.00 & 5.00 & 4.00 & 4.00 & 4.50 & 0.50 \\
    75 - 79 (N=8, 7.77\%) & 3.25 & 5.00 & 4.1875 & 4.75 & 5.00 & 0.8125 \\
    \hline
    \end{tabularx}
    \caption{Motivations of AI literacy education by age groups. The percentage for the number of participants is expressed in
relation to the total number of participants (N = 103).}
    \label{tab:motivations_across_age}
\end{table*}

\subsubsection{Motivation and Education}
\label{app:motivation-and-education}
We examined the relationship between participants' education level and their motivation for AI literacy education. Among the 103 participants, 2 indicated 'Prefer not to say' for their education level and thus were excluded from this analysis. Table \ref{tab:motivations_across_education} provides details for the remaining 101 participants, grouped by education level. The median motivation scores across all education levels are 4.00. Thus, the results suggest that the motivation to learn is not concentrated within any particular education level.

Also, we used Data Analysis ToolPak in Excel to conduct regression analysis between participants' education level and their motivation scores. We performed one-hot encoding for the education level variable, which is categorical. The Multiple R value was 0.12, indicating a weak or negligible correlation. The p-value associated with the F statistic is 0.78 which is greater than 0.05, implying that education does not have a statistically significant relationship with the motivation score.

 
                           



\begin{table*}
    \centering
    \small
    \renewcommand{\arraystretch}{1.6} 
    \begin{tabularx}{\textwidth}{p{3cm}p{2cm}p{2cm}p{2cm}p{2cm}p{2cm}p{2cm}}
    \hline
    \toprule
    \textbf{Education Level} & \textbf{Min} & \textbf{Max} & \textbf{25th Percentile} & \textbf{Median} & \textbf{75th Percentile} & \textbf{IQR} \\
    \hline
    Secondary school (N=18, 17.48\%) & 3.25 & 5.00 & 4.00 & 4.00 & 4.75 & 0.75 \\
    Tertiary education (N=27, 26.21\%) & 3.75 & 5.00 & 4.00 & 4.00 & 4.875 & 0.875 \\
    Bachelor's degree (N=42, 40.78\%) & 2.75 & 5.00 & 4.00 & 4.00 & 4.75 & 0.75 \\
    Master's degree (N=14, 13.59\%) & 3.25 & 5.00 & 3.8125 & 4.00 & 4.625 & 0.8125 \\
    \hline
    \end{tabularx}
    \caption{Motivations of AI literacy education by education level. The percentage for the number of participants is expressed in relation to the total number of participants (N = 103).}
    \label{tab:motivations_across_education}
\end{table*}

\subsubsection{Motivation and Other Demographic Characteristics}
\label{app:motivation-and-other-demographics}
We assessed the correlation between the motivation of participants for AI literacy education and other demographic characteristics, namely: Competency in digital literacy, readiness to accept technology, perceptions of AI and perceived self-efficacy of learning AI. We used Data Analysis ToolPak in Excel to perform a regression analysis examining the relationship between participants' demographic characteristics and their motivation scores. Table \ref{tab:correlation-motivation-demographic} presents the correlation coefficient, the p-value associated with the F statistic, and the corresponding conclusions based on the analysis.

\begin{table*}
    \centering
    \small
    \renewcommand{\arraystretch}{1.6} 
    \begin{tabular}{>{\columncolor{LightGray}}p{3cm}p{4cm}p{2cm}p{2cm} p{5cm}}
    \hline
    \textbf{Dependent Variable} & {\textbf{Independent Variable}} & {\textbf{Correlation\newline Coefficient}} & {\textbf{p-value\newline (F statistic)}} & \textbf{Conclusion} \\
    \hline

    Motivations of AI literacy education & Competency in digital literacy & 0.47 & 0.00000058 & p < 0.05. Statistically significant moderate positive relationship. \\
                           & Readiness to accept technology & 0.30 & 0.0025 & p < 0.05. Statistically significant weak to moderate positive relationship.\\
                           & Perceptions of AI & 0.41 & 0.000016 & p < 0.05. Statistically significant moderate positive relationship.\\
                           & Perceived self-efficacy of learning AI & 0.53 & 0.000000012 & p < 0.05. Statistically significant moderate positive relationship.\\
    \hline
\end{tabular}
  \caption{Regression analysis between participants' demographic characteristics and motivation scores for AI literacy education.}\label{tab:correlation-motivation-demographic}
\end{table*}